\documentclass[prl,aps,superscriptaddress,twocolumn,notitlepage,floatfix,10pt]{revtex4-2}

\usepackage{amsmath,mathtools,amsthm,amssymb,pifont}
\usepackage[utf8]{inputenc}
\usepackage[american]{babel}
\usepackage{graphicx,xcolor,bbold,titlesec}
\usepackage{braket}
\usepackage{MnSymbol}

\usepackage[colorlinks,
bookmarksopen,
bookmarksnumbered,
citecolor=red,
linkcolor=red,
pdfstartview=false,
urlcolor=red]{hyperref}
\usepackage{tikz,ifthen}
\usepackage{bbold}
\usepackage{tikz-network}
\usetikzlibrary{patterns,decorations.pathreplacing,calligraphy}
\usepackage{color}
\usepackage{MnSymbol}
\usepackage{orcidlink}

\definecolor{myyellow}{RGB}{240,188,66}
\definecolor{myorange}{RGB}{255,102,0}
\definecolor{myorangel}{RGB}{255,204,153}
\definecolor{myblue}{RGB}{66,135,245}

\newcommand{\be}{\begin{equation}}
\newcommand{\ee}{\end{equation}}
\newcommand{\bec}{\begin{equation*}}
\newcommand{\eec}{\end{equation*}}
\newcommand{\bea}{\begin{eqnarray}}
\newcommand{\eea}{\end{eqnarray}}

\newcommand{\titleinfo}{Anticoncentration and nonstabilizerness spreading under ergodic quantum dynamics}
 
\begin{document}
 \title{\titleinfo}

\author{Emanuele Tirrito~\orcidlink{0000-0001-7067-1203}}
\affiliation{The Abdus Salam International Centre for Theoretical Physics (ICTP), Strada Costiera 11, 34151 Trieste, Italy}
\affiliation{Dipartimento di Fisica ``E. Pancini", Universit\`a di Napoli ``Federico II'', Monte S. Angelo, 80126 Napoli, Italy}

\author{ 
Xhek Turkeshi~\orcidlink{0000-0003-1093-3771}}
\affiliation{Institut f\"ur Theoretische Physik, Universit\"at zu K\"oln, Z\"ulpicher Strasse 77, 50937 K\"oln, Germany}

\author{Piotr Sierant~\orcidlink{0000-0001-9219-7274}}
\affiliation{Barcelona Supercomputing Center, Barcelona 08034, Spain}

\date{\today}

\begin{abstract}
Quantum state complexity metrics, such as anticoncentration and nonstabilizerness, or ``magic'', offer key insights into many-body physics, information scrambling, and quantum computing.
Anticoncentration and equilibration of magic resources under dynamics of random quantum circuits occur at times 
scaling logarithmically with system size,
a prediction that is believed to extend to more general ergodic dynamics.  
This work challenges this idea by 
examining the anticoncentration and magic spreading in one-dimensional ergodic Floquet models and Hamiltonian systems. 
Using participation and stabilizer entropies to probe these resources, we reveal significant differences between the two settings. 
Floquet systems align with random circuit predictions, exhibiting anticoncentration and magic saturation at time scales logarithmic in system size. In contrast, 
Hamiltonian dynamics deviate from the random circuit predictions and require times scaling approximately linearly with system size to achieve saturation of participation and stabilizer entropies, which remain smaller than that of the typical quantum states even in the long-time limit.
Our findings establish the phenomenology of participation and stabilizer entropy growth in ergodic many-body systems and emphasize the role of conservation laws in constraining anticoncentration and magic dynamics.
\end{abstract}
\maketitle

\paragraph{Introduction. } 
Understanding the propagation of quantum complexity metrics, such as entanglement~\cite{horodecki2009quantum,amico2008entanglement}, anticoncentration~\cite{dalzell2022random}, and nonstabilizerness, or ``magic'',  resources~\cite{liu2022manybody}, offers foundational insights into conceptual challenges of many-body physics, including quantum chaos, information scrambling and the emergence of statistical mechanics~\cite{dalessio2016from,pappalardi2022eigenstate,abanin2019colloquium,sierant2024manybodylocalizationageclassical,Oliviero2022measuring}.
Addressing these problems for specific systems is notoriously difficult, as it requires unraveling the intricate structure of many-body states and computing non-linear measures such as entanglement~\cite{calabrese2009entanglement}, participation~\cite{luitz2014universal,mace2019multifractal,sierant2022universal}, and stabilizer entropies~\cite{leone2022stabilizerrenyientropy,leone2024stabilizer,haug2023quantifying}.
In this context, random quantum circuits have emerged as a powerful and analytically tractable framework for exploring quantum dynamics~\cite{fisher2023randomquantumcircuits,chan2018solution,shivam2023manybody,nahum2018operator}, bridging conceptual insights from many-body physics~\cite{piroli2020a,bertini2020scrambling,fava2024designsfreeprobability,deluca2023universality,claeys2021ergodic,dowling2023scrambling} to applications in quantum computing~\cite{boixo2018characterizing,arute2019quantum,wu2021strong}.
A cornerstone result has been understanding the propagation of entanglement~\cite{nahum2017quantum}, elegantly captured by the membrane picture~\cite{zhou2019emergent,zhou2020entanglementmembrane,vasseur2019entanglement,potter2022entanglement,sierant2023entanglement,sommers2024zero}.

More recently, random quantum circuits allowed us to resolve the evolution of anticoncentration and nonstabilizerness.
Anticoncentration, or Hilbert space delocalization, quantifies how much a state spreads in the computational basis~\cite{hangleiter2023computational}. 
Magic resources~\cite{bravyi2005universal,gross2006hudsonstheoremfor,Veitch2014theresourcetheory} measure how much a state deviates from stabilizer states~\cite{gottesman1998heisenbergrep,gottesmann1998faulttolerantquantum}. 
Within random quantum circuits, these properties exhibit similar phenomenology, reaching the stationary value of typical many-body states in a timescale logarithmic in the system size $t_\mathrm{sat}\propto \log_2(N)$~\cite{bertoni2024shallow,turkeshi2024hilbert,braccia2024computing,garciamartin2024architectures,turkeshi2024magic,christopoulos2024universaldistributionsoverlapsunitary, zhang2024quantumma,magni2025anticoncentrationcliffordcircuitsbeyond,magni2025quantumcomplexitychaosmanyqudit,zhang2025designsmagicaugmentedcliffordcircuits}.
The rapid saturation of anticoncentration and nonstabilizerness differ from the ballistic increase of the entanglement entropy, whose saturation timescale is linear in the system size $t_{\mathrm{sat}} \propto N$~\cite{Lauchli08, Kim13ballistic}.

\begin{figure}
    \centering
    \includegraphics[width=1\linewidth]{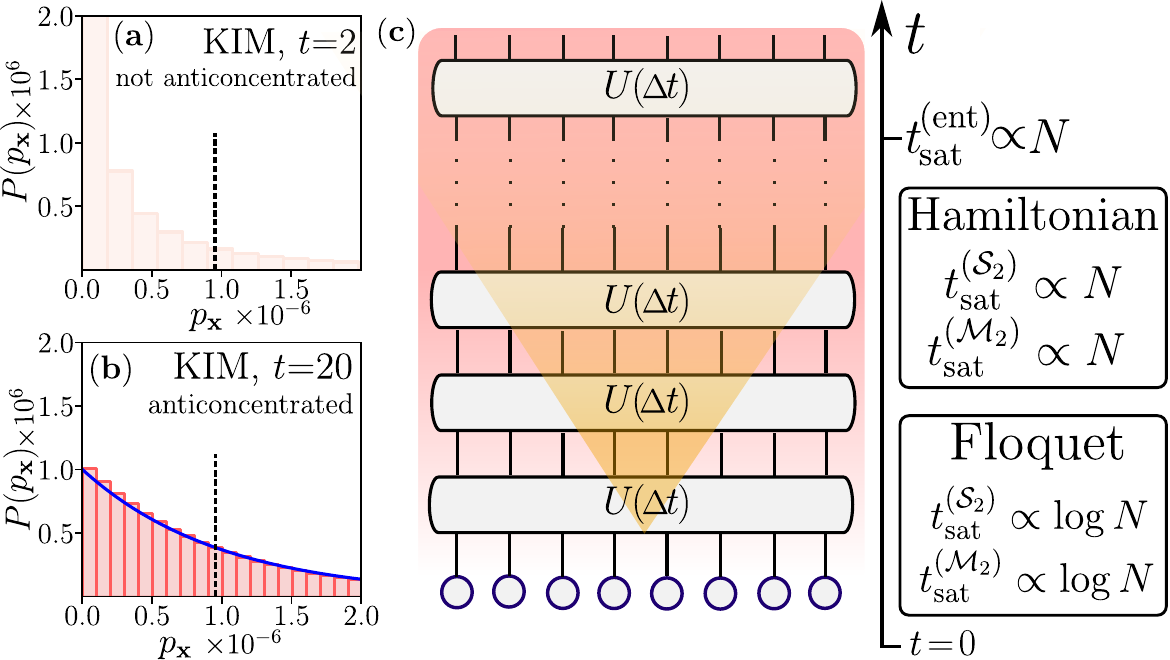}
    \caption{Anticoncentration and magic resources under the dynamics of a many-body system with time-evolution operator $U(\Delta t)$. The distribution of probabilities $p_{\pmb{x}}$ changes from being concentrated at small evolution time $t$, (a), to being anticoncentrated at large $t$, (b), and following the Porter-Thomas distribution~\cite{Porter56, mullane2020sampling} (blue line). (c) Entanglement, PE, $\mathcal{S}_2$, and magic resources (quantified by SE, $\mathcal{M}_2$) increase and reach, up to a fixed accuracy $\epsilon$ their saturation values at times $t^{(\mathrm{ent})}_{\mathrm{sat}}$, $t^{(\mathcal{S}_2)}_{\mathrm{sat}}$, $t^{(\mathcal{M}_2)}_{\mathrm{sat}}$. 
    }
    \label{fig:sketch}
\end{figure}

The success of random quantum circuits is tied with arguments of typicality, leading to the widespread belief that they capture the qualitative features of 
ergodic quantum systems. This work challenges that assumption by examining the dynamics of anticoncentration and nonstabilizerness in ergodic Floquet~\cite{moessner2017equilibration} and Hamiltonian systems~\cite{polkovnikov2011colloquium}.
Our findings reveal significant qualitative differences: while 
Floequet systems align with predictions from random circuits,
Hamiltonian systems exhibit markedly distinct dynamical behavior within accessible system sizes. 
We attribute these discrepancies to the 
presence of conservation laws, in particular energy conservation, which is intrinsic to Hamiltonian systems.
Specifically, we find that under Floquet dynamics, the participation entropy (\textbf{PE}) and the stabilizer entropy (\textbf{SE}) saturate, up to a fixed tolerance $\varepsilon <1$, to the Haar values on a timescale $t_\mathrm{sat}\propto \log_2(N)$. 
In contrast, for Hamiltonian dynamics, we observe a longer saturation time $t_{\mathrm{sat}}\propto N$, both for the PE and SE, see Fig.~\ref{fig:sketch}.

\paragraph{Quantities of interest.}
We study a quantum chain of $N$ qubits, with total Hilbert space dimension $D=2^N$. 
We will quantify anticoncentration with the PE, which measures the spread of a state over a basis $\mathcal{B}$.
For a many-body state $|\Psi\rangle$ the PE in the computational basis $\mathcal{B}_N=\{|\pmb{x}\rangle\}_{x=0}^{D-1}$ is 
\begin{equation} 
    \mathcal{S}_{k}(|\Psi\rangle) = \frac{1}{1-k}\log_2 \left[\sum_{\pmb{x}\in\mathcal{B}_N} p_{\pmb{x}}^k\right]\;,\quad p_{\pmb{x}}\equiv |\langle \pmb{x}|\Psi\rangle|^2\;,\label{eq:ipr}
\end{equation}
where $p_{\pmb{x}}$ is the probability to find the system in basis state $\ket{\pmb{x}}$.
Notice that 
$\mathcal{S}_{1}$ is defined by taking the limit $k \to 1$ in \eqref{eq:ipr} and that $\mathcal{S}_{k}\ge 0$, with equality holding if and only if $|\Psi\rangle$ is a 
state of the computational basis. 
In general, the system size dependence of the PE is~\cite{laflorencie2014spin,luitz2014improving,luitz2014shannon}
\begin{equation}
\label{eq:Ldep}
  \mathcal{S}_{k}(|\Psi\rangle)= D_k N +c_k,  
\end{equation}
where $D_k$ is the multifractal dimension and $c_k$ is the subleading term~\cite{backer2019multifractal,mace2019multifractal,sierant2022universal}. For $D_k=0$, the state $|\Psi\rangle$ is completely localized, while for $D_k=1$ the state is fully extended. In the intermediate case, $0<D_k<1$, when $D_k$ depends non trivially on the index $k$ 
the state is said to be multifractal.
The PE $\mathcal{S}_2$ is proportional to the logarithm of \textit{collision probability} $p_{\mathrm{col}}=\sum_{\pmb{x}\in\mathcal{B}_N} p_{\pmb{x}}^2$. The anticoncentration of state $\ket{\Psi}$ in basis $\mathcal{B}_N$, required by arguments of classical hardness of the sampling problems~\cite{aaronson2011computational, Bremner16, Bouland19, Oszmaniec22}, occurs~\cite{dalzell2022random} when $p_{\mathrm{col}} \leq a^{-1} 2^{-N}$, where $0 \leq a \leq 1$ is a constant. In terms of the PE scaling~\eqref{eq:Ldep}, the anticoncentration requires $D_k=1$ and $c_k \leq 0$ and can be verified on the level of the distribution of $p_{\pmb{x}}$, see Fig.~\ref{fig:sketch}~(a), (b).

We quantify nonstabilizerness via the SE, defined for a state $|\Psi\rangle$ in terms of the Pauli strings on $N$ qubits  $\mathcal{P}_N\equiv \{P_1\otimes P_2 \otimes \cdots \otimes P_N\}_{\{P_j=I,X,Y,Z\}}$ as~\cite{leone2022stabilizerrenyientropy}
\begin{equation}
    \mathcal{M}_{k}(|\Psi\rangle) = \frac{1}{1-k}\log_2\left[\sum_{P\in\mathcal{P}_N} D^{k-1}\Xi_P^k\right]\;,\quad \Xi_P = \frac{\langle \Psi | P |\Psi\rangle^2}{D}\;,\label{eq:sre}
\end{equation}
with $\Xi_P$ the Pauli spectrum~\cite{beverland2020lower,turkeshi2023paulispectrummagictypical}. 
Again, $\mathcal{M}_1$ is defined by the limit $k\to 1$ in~\eqref{eq:sre}, and $\mathcal{M}_{k}\ge 0$, with the equality holding if and only if $|\Psi\rangle$ is a stabilizer state~\cite{haug2023stabilizerentropiesand,gross2021schurweylduality}.  
A figure of merit of Eq.~\eqref{eq:sre} is its efficient computability for matrix product states~\cite{haug2023quantifying,tarabunga2024nonstabilizernessmatrixproductstates,lami2023nonstabilizernessviaperfect,tarabunga2023manybodymagic,tirrito2024quantifying,turkeshi2023measuring}.

Eqs.~\eqref{eq:ipr} and~\eqref{eq:sre} highlight a similarity between anticoncentration and nonstabilizerness spreading when $\mathcal{P}_N$ is viewed as a basis in the operator space. In particular, both PE and SE are basis-dependent quantities. Several connections between PE and SE have been found in recent works~\cite{haug2023quantifying,turkeshi2023measuring,haug2024efficient,turkeshi2024magic,collura2024quantummagicfermionicgaussian}.

\paragraph{Methods and models.}
Our main goal is to understand the growth of PE and SE under the dynamics of ergodic quantum many-body systems and, in particular, to determine the timescales at which PE and SE saturate to their long-time stationary values. 
We define $\Delta \mathcal{S}_{k}(t)= \mathcal{S}_{k}^\infty - \mathcal{S}_{k}(t)$, representing the deviation of the time-evolving participation entropy $\mathcal{S}_{k}(t)$ from the long-time saturation value $\mathcal{S}_{k}^\infty \equiv \lim_{t\to\infty} \mathcal{S}_{k}(t)$, as a measure of the Hilbert space delocalization. Similarly, we define $\Delta \mathcal{M}_{k}(t)= \mathcal{M}_{k}^\infty - \mathcal{M}_{k}(t)$ as the deviation of the time-evolving stabilizer entropy from its stationary value $\mathcal{M}_{k}^\infty  \equiv \lim_{t\to\infty} \mathcal{M}_{k}(t)$, serving as a measure for magic spreading. 
For local random one-dimensional unitary circuits of sufficient depth $t$, it is established~\cite{dalzell2022random, turkeshi2024hilbert, turkeshi2024magic} that
\begin{equation}
\Delta \mathcal{S}_k(t) = A_s e^{-\alpha_s t}, \quad \Delta \mathcal{M}_k(t) = A_m e^{-\alpha_m t},
   \label{eq:decay}
\end{equation}
where $\alpha_{s,m}$ are system-size independent constants, $A_{s,m}\propto N$. The relevant long-time saturation values of PE ($\mathcal{S}_{2}^\infty$) and SE ($\mathcal{M}_{2}^\infty$) are equal to the values for Haar-random states, which, for $k=2$, read  $ \mathcal{S}_{2}^{\mathrm{Haar}} = \log_2\left[D+1\right]-1$ and $\mathcal{M}_{2}^{\mathrm{Haar}}= \log_2[D+3]-2$~\cite{turkeshi2023paulispectrummagictypical, turkeshi2024hilbert}. Eq.~\eqref{eq:decay} implies that
$\Delta \mathcal{S}_{k}( t) \leq \varepsilon$ for $t>t^{(\mathcal{S}_k)}_{\mathrm{sat}}  \propto \log_2(N)$, showing that the anticoncentration, $p_{\mathrm{col}} \leq a^{-1} 2^{-N}$,  occurs on logarithmic timescales~\cite{dalzell2022random,turkeshi2024hilbert}. Similarly, magic spreading follows an analogous behavior, with $\Delta \mathcal{M}_{k}(t) \leq \varepsilon$ for $t>t^{(\mathcal{M}_k)}_{\mathrm{sat}}\propto \log_2(N)$~\cite{turkeshi2024magic}.
The results of random quantum circuits will serve as a reference point in our investigation of the time evolution of PE and SE under the dynamics of ergodic quantum many-body systems~\cite{dalessio2016from}.

\begin{figure*}
    \centering
    \includegraphics[width=1\linewidth]{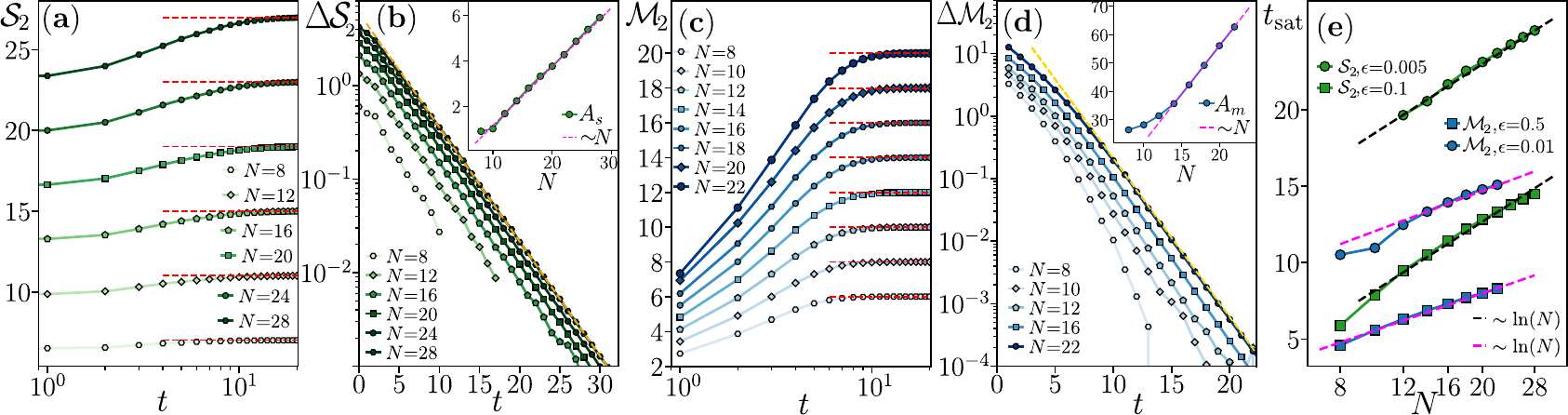}
    \caption{ 
    Anticoncentration and magic spreading in KIM for $N$ qubits.
    (a) PE, $\mathcal{S}^{(2)}$, saturates to $S^{\mathrm{Haar}}_2$ (dashed line). (b) An exponential  decay of $\Delta \mathcal{S}_2 = A_s e^{-\alpha_S t}$ with time $t$, where $\alpha_S=0.28(2)$, and $A_s \propto N$, see the inset.
    (c) the SE, $\mathcal{M}_2$, abruptly saturates to $\mathcal{M}^{\mathrm{Haar}}_2$ (dashed line).
    (d) The difference $\Delta \mathcal{M}_2$ is well fitted by an  exponential decay, $\Delta \mathcal{M}_2 = A_m e^{-\alpha_M t}$, with $\alpha_M=0.59(3)$, and $A_m \propto N$, see the inset.
    (e) $\Delta \mathcal{S}_2$ and $\Delta \mathcal{M}_2$ decay to a given  $\epsilon \lesssim O(1)$ at times $t_{\mathrm{sat}} \propto \log_2(N)$ scaling logarithmic with system size $N$.The results are averaged over more than $200$ initial states $\ket{\Psi_0}$~\eqref{eq:initial_state}.
    }
    \label{fig:floquet_circuit}
\end{figure*}

We consider two paradigmatic ergodic quantum many-body systems: a Kicked Ising model (KIM)~\cite{Prosen98, Prosen99, Prosen02}, with Floquet operator defined by 
\begin{equation}
    U_\mathrm{KIM}= e^{-i b \sum_j X_j} e^{-i \sum_j h_j Z_j + J\sum_j Z_j Z_{j+1}}\;.\label{eq:kimm}
\end{equation}
and a mixed fields Ising model (MFIM)~\cite{Kim13ballistic} with Hamiltonian given by
\begin{equation}
    \label{eq:Ising_Hamiltonian}
H_{\mathrm{MFIM}}=b \sum_j X_j + \sum_j h_j Z_j + J\sum_j Z_j Z_{j+1}  \;,
\end{equation} 
where $b, h_j, J$ are parameters, which we fix as $b = (\sqrt{5}+5)/8$, $h_i = (\sqrt{5}+1)/4$, $J=1$, following the choice of~\cite{Kim13ballistic}, and assume open boundary conditions for both models. The trends and conclusions of our investigations are expected to not depend on the precise choice of $b, h_j, J$, as long as the parameters are not fine-tuned to special points such as the dual-unitarity point of KIM (when $|b| = |J| = \pi/4$~\cite{Akila16, Bertini18, Bertini19dual, bertini2022entanglement,rampp2023from,foligno2024quantum}) or free-fermionic point of MFIM (for $h_j=0$)~\footnote{In the dual-unitary limit, the evolution of PE has been studied in Ref.~\cite{claeys2024fockspacedelocalizationemergenceporterthomas} while that of ME in Ref.~\cite{kos2024exact,dowling2024magicheisenbergpicture}. The magic spreading under free fermionic Hamiltonian evolution has been studied in Ref.~\cite{rattacaso2023stabilizer}. }.
We consider an initial state of the form 
\begin{equation}
    \label{eq:initial_state}
|\Psi_0 \rangle= \prod_{j=1}^N U^{(1)}_j \ket{0}^{\otimes N},
\end{equation}  
where $U^{(1)}_j$ are independent Haar-random unitaries acting on site $j$. We compute time-evolved state $|\Psi_t\rangle = U_\mathrm{KIM}^t |\Psi_0\rangle$ for KIM employing fast Hadamard transform~\cite{Lezama19, Sierant23flo}, and $|\Psi_t\rangle = e^{-i H_{\mathrm{MFIM}}t} |\Psi_0\rangle$ using Chebyshev time-evolution~\cite{Tal‐Ezer84, Sierant22chal}.
We calculate PE directly from the definition~\eqref{eq:ipr} and employ the algorithm of~\cite{sieranttoappear} to calculate SE. To obtain insights into the generic behavior of PE and SE, we average $\mathcal{S}_{2}(t)$ and $\mathcal{M}_{2}(t)$ over more than $200$ choices of the initial state $|\Psi_0 \rangle$.

\paragraph{Results for Floquet dynamics.}
We first study the anticoncentration and magic spreading in the Floquet dynamics of KIM. 
\begin{figure*}
    \centering
    \includegraphics[width=1\linewidth]{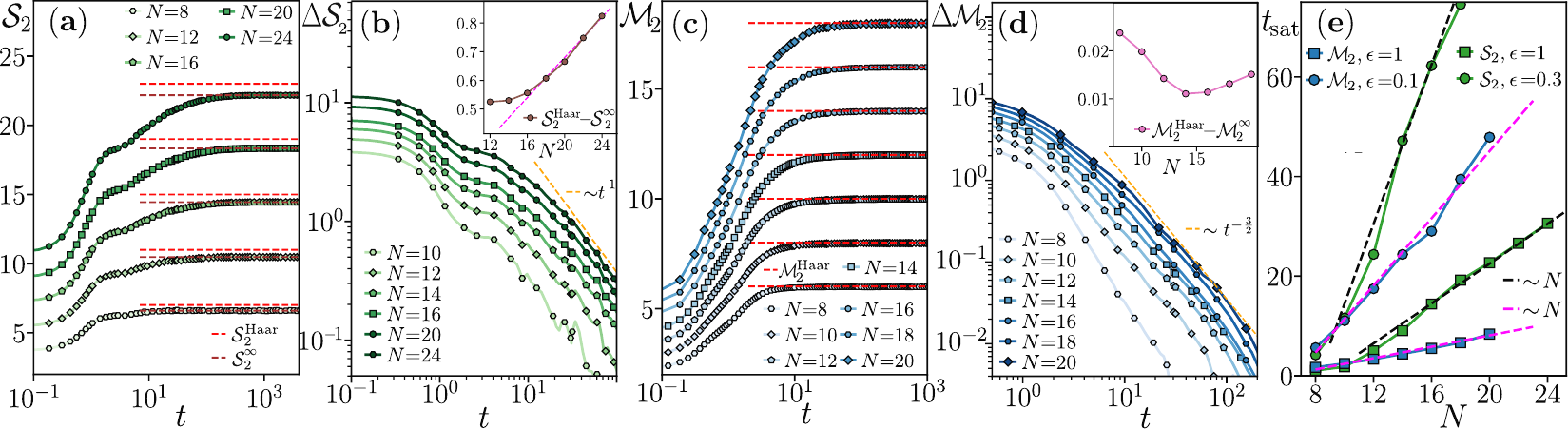}
    \caption{
    Anticoncentration and magic spreading in MFIM for $N$ qubits.
    (a) The PE, $\mathcal{S}^{(2)}$, increases towards 
    the long-time saturation value $\mathcal{S}^{\infty}_2$ whose difference with $\mathcal{S}_2^{\mathrm{Haar}}$ remains non-zero, see the inset in (b). The difference $\Delta \mathcal{S}_2$, see panel (b),  follows a power-law decay, $\Delta \mathcal{S}_2 \propto t^{-
\beta_S}$ with $\beta_S \approx 1$.
    (c) The SE, $\mathcal{M}_2$, saturates to $\mathcal{M}^{\infty}_2$ smaller than the Haar value $\mathcal{M}^{\mathrm{Haar}}_2$, as shown in the inset in (d).
    The difference $\Delta \mathcal{M}_2$, shown in (d), is well fitted by a power-law decay $\Delta \mathcal{M}_2  \propto t^{-\beta_M}$ with $\beta_M \approx -1.5$.
    (e) $\Delta \mathcal{S}_2$ and $\Delta \mathcal{M}_2(t)$ decay to a given $\epsilon $ at times $t_{\mathrm{sat}} \propto N$ scaling linearly with system size $N$.
    }
    \label{fig:MFIM}
\end{figure*}
Our results are summarized in Fig.~\ref{fig:floquet_circuit}.
We observe that PE and SE saturate for $t \gg 1$ to the values characteristic of Haar-random states, $\mathcal{S}_{k}^\infty = \mathcal{S}_{k}^{\mathrm{Haar}}$, and $\mathcal{M}_{k}^\infty= \mathcal{M}_{k}^{\mathrm{Haar}}$, as shown in Fig.~\ref{fig:floquet_circuit} (a) and (c). 
Moreover, as shown in Fig.~\ref{fig:floquet_circuit}~(b), for $t \gtrapprox 5$, the difference $\Delta \mathcal{S}_{2}(t)=\mathcal{S}_{2}^{\mathrm{Haar}}- \mathcal{S}_{2}(t)$ decays exponentially in time  as
$\Delta \mathcal{S}_{2} = A_s e^{-\alpha_s t}$,
where $\alpha_s = 0.28(2)$ is, at sufficiently large system size, independent of $N$. 
The parameter $A_s$ scales linearly with $N$ for $N \gtrsim 12$, see the inset in Fig.~\ref{fig:floquet_circuit}~(b). 
The extended range of system sizes $N \le 28$ accessible to our computations allows us to observe a logarithmic scaling, $\tau_\mathrm{sat}^{(\mathcal{S}_2)} \propto\log_2(N)$, of the time $\tau_\mathrm{sat}^{(\mathcal{S}_2)}$ beyond which $\Delta \mathcal{S}_k < \epsilon$, see Fig.~\ref{fig:floquet_circuit} (e). 
The phenomenology of the PE growth and its saturation under dynamics of KIM fully parallels the random circuit models behavior~\cite{turkeshi2024hilbert}.

The evolution of SE, $\mathcal{M}_2$, as a function of time $t$ for system sizes $N\le 22$ is shown in Fig.~\ref{fig:floquet_circuit}~(c); similar results hold for other R\'enyi indices $k$ (see~\cite{supmat}). We observe a transient regime in which $\mathcal{M}_2$ increases rapidly before saturating to the Haar value $\mathcal{M}^{\mathrm{Haar}}_2$.  
In Fig.~\ref{fig:floquet_circuit}~(d), we 
observe that $\Delta \mathcal{M}_2$ decays exponentially as a function of time $t$, and the prefactor $A_m$ is proportional to the system size $N$.
Moreover, Fig.~\ref{fig:floquet_circuit}~(e) shows the saturation time $t^{(\mathcal{M}_{2})}_{\mathrm{sat}}$ of SE as a function of system size $N$, 
revealing a logarithmic scaling of $t^{(\mathcal{M}_{2})}_{\mathrm{sat}}$ with system size,  analogous to the behavior in local random quantum circuits~\cite{turkeshi2024magic}.

\paragraph{Results for Hamiltonian dynamics.}
We now turn to non-integrable Hamiltonian dynamics, focusing on the MFIM.
For the parameters considered, the model is quantum ergodic~\cite{Kim13ballistic}, and its level statistics follow the predictions of Gaussian Orthogonal Ensemble of random matrices~\cite{Mehtabook,Haakebook}.  
We consider initial states $|\Psi_0\rangle$ of the form~\eqref{eq:initial_state}, with additional constraint that $ |\bra{\Psi_0 } H_{\mathrm{MFIM}} \ket{\Psi_0 }-E_{\mathrm{min}}|/(E_{\mathrm{max}}-E_{\mathrm{min}}) \leq 0.05$,
ensuring that the initial state energy lies close to the middle of the spectrum. Here $E_{\mathrm{max}}$ and $E_{\mathrm{min}}$ are respectively the maximal and minimal eigenvalues of $H_{\mathrm{MFIM}}$.

Our results for PE and SE growth under MFIM dynamics are summarized in Fig.~\ref{fig:MFIM}. 
The PE, $\mathcal{S}_{2}$, is shown as a function of time $t$ in Fig.~\ref{fig:MFIM}~(a). Already at time $t=1$, $\mathcal{S}_{2}$ is proportional to the system size $N$. At later times, the PE saturates to a stationary value $\mathcal{S}^{\infty}_2$ that is close to—but distinct from—the Haar value $\mathcal{S}^{\mathrm{Haar}}_2$. This marks the first qualitative difference between MFIM dynamics and the behavior observed in  Floquet models.  
In Fig.~\ref{fig:MFIM}~(b), we present the difference $\Delta \mathcal{S}_2(t) = \mathcal{S}^{\infty}_2 - \mathcal{S}_2(t)$ for various system sizes, where $\mathcal{S}^{\infty}_2$ is computed by averaging $\mathcal{S}_2(t)$ over time interval $t\in[1000,5000]$. For $t \gtrapprox 5$, $\Delta \mathcal{S}_{2}$ 
decays as a power-law in time,
\begin{equation}
\Delta \mathcal{S}_2(t) = a_d t^{-\beta_S},
\label{eq:Stfim}
\end{equation}
where $\beta_S \approx 1.0$, although the accessible range of system sizes and time scales prevents us from pin-pointing the value of $\beta_S$. The algebraic decay~\eqref{eq:Stfim} differs qualitatively from the exponential decay observed for Floquet dynamics.
The inset in Fig.~\ref{fig:MFIM}~(b) shows the difference between the Haar value $\mathcal{S}^{\mathrm{Haar}}_2$ and the long-time saturation value $\mathcal{S}^{\infty}_2$ under MFIM dynamics.
We find that $\mathcal{S}^{\mathrm{Haar}}_2-\mathcal{S}^{\infty}_2$ starts to increase linearly with system size for $N\geq 16$, which implies that $D_2 < 1$ in \eqref{eq:Ldep}. Consequently, $p_{\mathrm{col}} > 2^{-N}$ at sufficiently large $N$, indicating that full anticoncentration of the time-evolved state $\ket{\Psi_t}$ in the computational basis $\mathcal{B}_N$ is not achieved, at least at the time scales and system sizes available to our numerical simulations.

In Fig.~\ref{fig:MFIM} (c), we show the evolution of the SE under MFIM dynamics. 
The initial state $\ket{\Psi_0}$ has a non-vanishing SE.  
Our data reveal a transient regime during which $\mathcal{M}_{2}(t)$ grows more slowly than in the KIM case, before saturating to the long-time value $\mathcal{M}^{\infty}_2$ that is close to —but noticeably smaller than—the Haar value $\mathcal{M}^{\mathrm{Haar}}_2$ (see the inset in Fig.~\ref{fig:MFIM}~(d)). 
For $t \gtrsim 1$, the difference $\Delta \mathcal{M}_2(t) = \mathcal{M}^{\infty}_2 - \mathcal{M}_2(t)$ increases with the system size $N$ and decays as a power-law in time:
\begin{equation}
\Delta \mathcal{M}_2(t) = a_d t^{-\beta_M},
   \label{eq:decayMFIM}
\end{equation}
where $a_d$ and $\beta_M\approx 1.5$ are constants. The power-law relaxation of the SE to its long-time value $\mathcal{M}^{\infty}_2$ under MFIM dynamics is one of the main results of this work.  
The algebraic decay of both $\Delta \mathcal{S}_2$ and $\Delta \mathcal{M}_2$ implies that the saturation of PE and SE occurs at significantly longer timescales than in the KIM. 
In Fig.~\ref{fig:MFIM}~(e), we show the the saturation times as a function of $N$ for different thresholds $\epsilon$. 
Both $t^{(\mathcal{S}_{2})}_{\mathrm{sat}}$ and $t^{(\mathcal{M}_{2})}_{\mathrm{sat}}$ 
scale approximately linearly with the system size $N$,  indicating qualitatively slower dynamics than those observed in Floquet models.

\paragraph{Discussion.}
The striking contrast between the rapid PE and SE saturation in the KIM, at timescales proportional to $\ln(N)$, and the much slower saturation in the MFIM, with timescales linear in $N$, reveals a fundamental difference between the Hamiltonian and Floquet dynamics.
This qualitative distinction persists even in the presence of weak disorder~\cite{supmat}, underscoring the robustness of the underlying mechanisms.

To understand the origin of this discrepancy, we note that energy conservation in the MFIM is the key feature that distinguishes it from the KIM.
Moreover, in the End Matter, we show that $U(1)$-\emph{symmetric} random quantum circuits behave similarly to the MFIM, exhibiting power-law saturation of PE and SE, with characteristic timescales that scale extensively with system size $N$—in contrast to the exponential behavior described by Eq.~\eqref{eq:decay}, which applies to circuits \emph{without} symmetries.
We further show that PE and SE dynamics in Floquet models follow Eq.~\eqref{eq:decay}, regardless of whether the Floquet operator can be implemented as a shallow circuit. The slower, power-law saturation described by Eqs.~\eqref{eq:Stfim} and~\eqref{eq:decayMFIM} emerges only when energy conservation is restored.

In view of this evidence, we propose that the mechanism responsible for the difference between fast ($t^{(\mathcal{S}_2, \mathcal{M}2)}{\mathrm{sat}} \propto \ln(N)$) and slow ($t^{(\mathcal{S}_2, \mathcal{M}2)}{\mathrm{sat}} \propto N$) saturation of PE and SE (cf. Fig.\ref{fig:sketch}) critically relies on the presence of conservation laws associated with operators that are sums of local quantities. When such a conservation law is present, local dynamics require a time that scales at least linearly with system size $N$ to fully redistribute the conserved charges. Only after this relaxation process completes do PE and SE saturate to their long-time values, $\mathcal{S}_2^{\infty}$ and $\mathcal{M}_2^{\infty}$.
This mechanism is consistent with recent arguments showing that symmetric local circuits cannot approximate symmetric Haar-random unitaries unless the circuit depth scales at least linearly with system size~\cite{haah2025short}.
Finally, the presence of atypical, area-law entangled low-lying eigenstates~\cite{Eisert10area,supmat}, which constrain the manifold of states explored by $\ket{\Psi_t}$ during its time evolution, prevents PE and SE from reaching the values characteristic of typical Haar-random states, $\mathcal{S}^{\mathrm{Haar}}_2$ and $\mathcal{M}^{\mathrm{Haar}}_2$, under MFIM dynamics.

\paragraph{Conclusion.}
In this work, we studied the anticoncentration and magic resources growth in the dynamics of ergodic quantum many-body systems. By using PE to characterize anticoncentration and SE to track the magic spreading, we revealed the essential role of conservation laws in shaping these processes, identifying qualitative differences between ergodic Floquet and Hamiltonian dynamics. While Floquet dynamics leads to the saturation of PE and SE at characteristic timescales that grow logarithmically with system size $N$, energy conservation in Hamiltonian dynamics delays equilibration, resulting in saturation only at times that scale extensively with $N$.

Two striking consequences emerge from our findings. First, simulating anticoncentration and magic growth in many-body dynamics is computationally more efficient for quantum circuits than for Hamiltonian dynamics. Digital, e.g., Floquet systems saturate faster, allowing tensor network methods~\cite{Schollw_ck_2011, Ors2019, Ran20} to efficiently describe their evolution with polynomial bond dimensions. 
This has implications for approximate shadow tomography protocols that combine tensor networks with shallow circuits are significantly more efficient for digital systems than for Hamiltonian ones~\cite{Huang_2020,Huang2022,mcginley2023shadow,heinrich2023randomizedbenchmarkingrandomquantum,brieger2023stabilityclassicalshadowsgatedependent, fux2024disentanglingunitarydynamicsclassically}. This also motivates further explorations of Clifford-augmented matrix product states~\cite{masotllima2024stabilizertensornetworks, Qian24prl, lami2024quantumstatedesignsclifford,qian2024cliffordcircuitsaugmentedtimedependent, nakhl2024sta} in scenarios where the dynamics are governed by continuous Hamiltonians.

Our findings open several avenues for future research. Energy conservation plays a major role in the anticoncentration and magic spreading.  A systematic investigation of other conservation laws, such as continuous symmetries, is essential. This may include analyzing the role of non-abelian symmetries and providing links between magic resources and the system’s slow hydrodynamic modes~\cite{khemani2018operator,curt2018operator,tibor2019sub, Michailidis24Corrections}. Furthermore, the impact of integrability~\cite{Vidmar_2016} and ergodicity breaking~\cite{abanin2019colloquium, Serbyn21, sierant2024manybodylocalizationageclassical}, which provide additional constraints on the manifolds of states visited throughout the time evolution, on anticoncentration and magic spreading remains unresolved. 
These questions are left for future work.

\begin{acknowledgments}
\paragraph{Acknowledgements.}
We thank L. Leone, L. Piroli, T. Haug, M. Heinrich, for discussions, and J. De Nardis, G. Lami, G. Fux, P. Tarabunga, M. Frau, N. Dowling, P. Kos, M. Dalmonte, R. Fazio, M. Savage, C. Robin, P. R. Nicácio Falcão, P. Stornati, S. Masot-Llima, A. Garcia-Saez for collaborations on related topics. 
E.\,T. acknowledges support from  ERC under grant agreement n.101053159 (RAVE), and
CINECA (Consorzio Interuniversitario per il Calcolo Automatico) award, under the ISCRA 
initiative and Leonardo early access program, for the availability of high-performance computing resources and support.
P.S. acknowledges fellowship within the “Generación D” initiative, Red.es, Ministerio para la Transformación Digital y de la Función Pública, for talent attraction (C005/24-ED CV1), funded by the European Union NextGenerationEU funds, through PRTR.
X.T. acknowledges support from DFG under Germany's Excellence Strategy – Cluster of Excellence Matter and Light for Quantum Computing (ML4Q) EXC 2004/1 – 390534769, and DFG Collaborative Research Center (CRC) 183 Project No. 277101999 - project B01.

\textit{Data Availability. ---}
Numerical data and snippets of the code are available in Ref.~\cite{dataavail}.

\textit{Note Added. ---}
During the completion of this work, we become aware of a complementary study by the group of A. Hamma. This work will appear in the same arXiv posting~\cite{hammatoappear}.
\end{acknowledgments}

\newpage

\section{End Matter.}

In this work, we explore the dynamics of Hilbert space delocalization and magic spreading under unitary Hamiltonian evolution. We focus on the KIM and MFIM as two paradigmatic models of ergodic many-body dynamics, and uncover striking differences in the growth of anticoncentration and nonstabilizerness in these systems. 
We note that the averaged $\mathcal{S}_2(t)$ and $\mathcal{M}_2(t)$ represent a typical time-evolution of PE and SE, and there is a non-negligible spread of $\mathcal{S}_2(t)$ and $\mathcal{M}_2(t)$ corresponding to the choice of the initial state $\ket{\Psi_0}$ for the dynamics of MFIM, discussed further in~\cite{supmat}. 
In the following, we analyze the role of the system's symmetry on the time evolution of the PE and SE.

\paragraph{Anticoncentration and nonstabilizerness growth in $U(1)$ symmetric circuits.}
Random quantum circuits without any symmetries provide the baseline prediction for saturation of PE and SE on time scales logarithmic in system size $N$, as expressed in Eq.~\eqref{eq:decay}. In the Main Text, we showed that the growth of PE and SE under KIM dynamics follows the random circuit predictions, and argued that energy conservation in the MFIM leads to PE and SE saturation at much longer times scaling extensively with $N$.

In the following, we investigate PE and SE growth in brick-wall $U(1)$ symmetric quantum circuits acting on $N$ qubits, with the evolution operator $U_t = \prod_{r=1}^t U^{(r)}$, where $t$ is the circuit depth (or time) and
\begin{equation}
    U^{(2m)} = \prod_{i=1}^{N/2-1} U_{2i,2i+1}\;,\quad U^{(2m+1)} = \prod_{i=1}^{N/2} U_{2i-1,2i}\;.
    \label{eq:randC}
\end{equation}
Here, the two-qubit gates $U_{i,j}$ acting on qubits $i$ and $j$ are independently drawn from the Haar measure on the subgroup of the unitary group $\mathcal{U}(4)$ that conserves total magnetization, i.e., $[U_{i,j}, \sum_l Z_l] = 0$, where $[.,.]$ denotes the commutator. The initial state is chosen as a product state of eigenstates of the $Z_j$ operators: $\ket{\Psi_0} = \ket{\sigma_1 \ldots \sigma_N}$. The spin values $\sigma_i \in {0,1}$ are selected randomly, subject to the constraint $\sum_j \sigma_j = N/2$, or equivalently, $\sum_l Z_l \ket{\Psi_0} = 0$—i.e., $\ket{\Psi_0}$ lies in the zero total magnetization sector.

\begin{figure}
    \centering
    \includegraphics[width=1\linewidth]{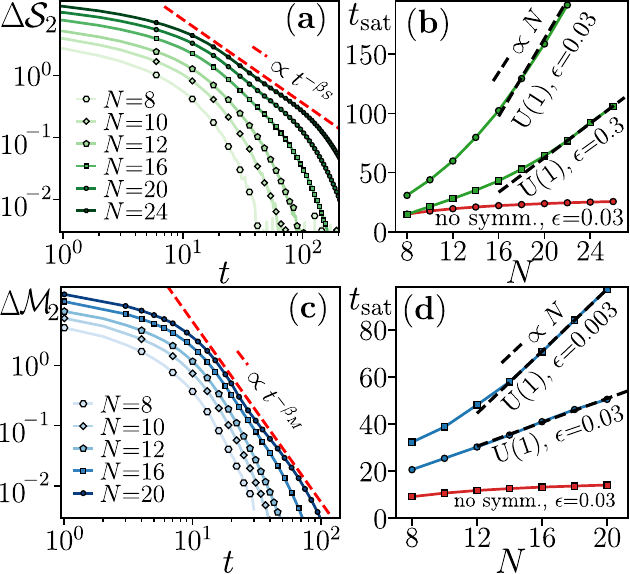}
    \caption{
Anticoncentration and magic dynamics in random unitary circuits with $U(1)$ symmetry.
(a), (c) Algebraic decay of the differences $\Delta \mathcal{S}_2 \propto t^{-\beta_S}$ and $\Delta \mathcal{M}_2 \propto t^{-\beta_M}$, with $\beta_S \approx 1.25$ and $\beta_M \approx 3$.
(b), (d) The saturation time scales extensively, $t_{\mathrm{sat}} \propto N$, for $U(1)$-symmetric circuits—in stark contrast to the $t_{\mathrm{sat}} \propto \ln(N)$ behavior observed in Haar-random brick-wall circuits without symmetry (red lines).}
    \label{fig:u1EM1}
\end{figure}
Under the dynamics of $U(1)$-symmetric random circuits, the time-evolved state $\ket{\Psi_t} = U_t \ket{\Psi_0}$ remains in the zero total magnetization sector, i.e., $\sum_l Z_l \ket{\Psi_t} = 0$. The presence of $U(1)$ symmetry dramatically alters the dynamics of PE and SE compared to random circuits without symmetries~\cite{turkeshi2024hilbert, turkeshi2024magic}.
As shown in Fig.~\ref{fig:u1EM1}, the differences $\Delta \mathcal{S}_2 = \mathcal{S}_2^{\infty} - \mathcal{S}_2(t)$ and $\Delta \mathcal{M}_2 = \mathcal{M}_2^{\infty} - \mathcal{M}_2(t)$—where $\mathcal{S}_2^{\infty}$ and $\mathcal{M}_2^{\infty}$ denote the long-time saturation values of PE and SE—decay algebraically in time: $\Delta \mathcal{S}_2 \propto t^{-\beta_S}$ and $\Delta \mathcal{M}2 \propto t^{-\beta_M}$.
This behavior stands in stark contrast to that of random circuits without symmetries, which exhibit exponential decay as described by Eq.~\eqref{eq:decay}. Consequently, the saturation times of PE and SE up to a fixed accuracy $\epsilon$ in $U(1)$-symmetric circuits scale extensively: $t^{(\mathcal{S}2)}{\mathrm{sat}} \propto N$ and $t^{(\mathcal{M}2)}{\mathrm{sat}} \propto N$, in contrast to the much faster saturation $t{\mathrm{sat}} \propto \ln(N)$ observed in circuits without symmetry.

These two features of PE and SE dynamics in $U(1)$-symmetric circuits closely mirror the phenomenology observed in Hamiltonian systems with energy conservation. This confirms that the key mechanism responsible for the slowdown in PE and SE growth is the presence of conservation laws. The relevant conserved quantity may be energy, as in Hamiltonian systems, or the $U(1)$ charge in symmetric circuits. In both cases, the conserved operator takes the form of a sum of local charges.
In contrast, we have verified (data not shown) that brick-wall circuits with only a $Z_2$ symmetry—i.e., one not associated with a locally conserved density—exhibit PE and SE growth consistent with Eq.~\eqref{eq:decay}, as in systems without conservation laws.

\paragraph{Family of Floquet models.}
The KIM, \eqref{eq:kimm}, can be recast as a two-layer quantum circuit containing local gates. From this perspective, adherence of the results for PE and SE growth in KIM to the predictions for random quantum circuits, \eqref{eq:decay} may be expected. In the following, we investigate dynamics under Floquet operator that cannot be written as a shallow circuit. 

We considered a family of models indexed by a parameter $\theta$ and defined by the following evolution operator:
\begin{equation}
    U_F=e^{-i\left[ \theta H_x+(1-\theta)H_z \right]} e^{-i\left[(1-\theta)H_x +\theta H_z\right]}
    \label{eq:UFdef}
\end{equation}
where $H_x= b\sum_j X_j$ and $H_z=J \sum_j  Z_j Z_{j+1} +\sum_j   h_j Z_j$ with $b = (\sqrt{5}+5)/8$, $h_i = (\sqrt{5}+1)/4$, $J=1$, following the choice of the parameters of the Main Text. The evolution operator $U_F$ defines a family models, which interpolate between the KIM at $\theta=0$ (when $U_F=U_{\mathrm{KIM}}$) and the MFIM at $\theta=1/2$ (when $U_F=e^{iH_{\mathrm{MFIM}}}$).
For intermediate values of $\theta \in (0, 1/2)$, the dynamics smoothly interpolate between the KIM and the MFIM. Importantly, the dynamics under $U_F$ does \textit{not} conserve energy for any $0 \leq \theta < 1/2$. Moreover, $U_F$, for any $0 < \theta < 1/2$, cannot be decomposed into a shallow quantum circuit composed of local gates.
Hence, the family of models~\eqref{eq:UFdef} is ideally suited for testing the impact of energy conservation and the presence (or absence) of circuit structure of the evolution operator on the dynamics of PE and SE.

\begin{figure}
    \centering
    \includegraphics[width=1\linewidth]{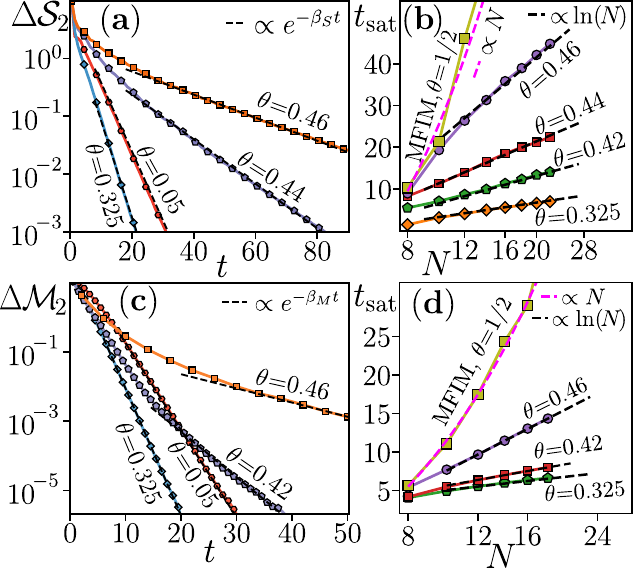}
    \caption{
Anticoncentration and magic dynamics in Floquet model~\eqref{eq:UFdef}.
(a), (c) The exponential decay of the differences $\Delta \mathcal{S}_2 \propto e^{-\beta_S t}$ and $\Delta \mathcal{M}_2 \propto e^{-\beta_M t}$ for system size $N=20$ and $0<\theta<1/2$. The decay slows down as $\theta$ approaches $1/2$.
(b), (d) The saturation times follow $t_{\mathrm{sat}} \propto \ln(N)$ for any $0<\theta<1/2$ ($\epsilon=0.2$ for PE and $\epsilon=0.1$ for SE) and scale extensively only for $\theta=1/2$, i.e., when the energy is conserved.
}
    \label{fig:floqEM1}
\end{figure}

Our results are shown in Fig.~\ref{fig:floqEM1}. The phenomenology of the growth of PE and SE under $U_F$ for any $0 < \theta < 1/2$ closely resembles the findings for the KIM reported in the Main Text, with the exponential relaxation of PE and SE towards the saturation values that coincide with those characteristic of Haar-random states. 
We also observe a non-monotonic dependence of the relaxation rate on the value of $\theta$: saturation at $\theta = 0.325$ occurs faster than at $\theta = 0.05$ and in the $\theta \to 0$ limit. Notably, for any $0 \leq \theta < 0.5$, the saturation of PE and SE to their respective Haar-state values (up to a fixed accuracy $\epsilon$) occurs at a time $t_{\mathrm{sat}}$ that scales logarithmically with the system size $N$. This matches our findings for the KIM reported in the Main Text, even though the operator $U_F$ does not admit a decomposition in terms of a shallow local circuit for any $0 < \theta < 1/2$.
The relaxation rate becomes slower as $\theta$ approaches the MFIM limit ($\theta = 1/2$), but even for $\theta = 0.46$, we observe that the dynamics exhibit exponential relaxation of PE and SE toward their saturation values. A qualitative change in behavior occurs only at $\theta = 1/2$, where the dynamics reduce to those of the MFIM and energy conservation is restored. 

The findings for the family of models $U_F$ defined by Eq.~\eqref{eq:UFdef} support the conclusion that energy conservation is the key factor responsible for the differences in the dynamics of PE and SE between the KIM and MFIM.

%

\newpage 

\widetext
\begin{center}
\textbf{\large \centering Supplemental Material:\\ Anticoncentration and nonstabilizerness spreading under ergodic quantum dynamics}
\end{center}

\setcounter{equation}{0}
\setcounter{figure}{0}
\setcounter{table}{0}
\setcounter{page}{1}
\renewcommand{\theequation}{S\arabic{equation}}
\setcounter{figure}{0}
\renewcommand{\thefigure}{S\arabic{figure}}
\renewcommand{\thepage}{S\arabic{page}}
\renewcommand{\thesection}{S\arabic{section}}
\renewcommand{\thetable}{S\arabic{table}}
\makeatletter

\renewcommand{\thesection}{\arabic{section}}
\renewcommand{\thesubsection}{\thesection.\arabic{subsection}}
\renewcommand{\thesubsubsection}{\thesubsection.\arabic{subsubsection}}

\vspace{0.3cm}
The Supplemental Material contains additional results supporting the conclusions about the phenomenology of PE and SE growth in ergodic many-body systems. In the following, we discuss:
\begin{itemize}
    \item Initial state dependence of the PE and SE growth in KIM and MFIM 
    \item Robustness of the PE and SE growth phenomenology upon introduction of disorder 
    \item The PE and SE in highly excited eigenstates of ergodic many-body systems
    \item Dynamics of PE and SE for R\'{e}nyi indices $k=1,3$.
\end{itemize}

\section{Initial state dependence}

As shown in the Main Text, in the case of KIM, both PE and SE rapidly equilibrate to their long-time saturation values, which coincide with those of random Haar states. 
The saturation of PE and SE follows
\be \label{eq:PE_SE_KIM}
\mathcal{S}_2=\mathcal{S}^{\mathrm{Haar}}_2-A_s e^{-\alpha_S t} \quad \mathrm{and} \quad \mathcal{M}_2=\mathcal{M}^{\mathrm{Haar}}_2-A_m e^{-\alpha_M t}
\ee
In other words, Eqs.~\eqref{eq:PE_SE_KIM} imply that the PE and the SE approach their stationary value, up to a fixed accuracy, at a time scaling logarithmically with the system size $N$. 

In contrast, MFIM displays a qualitatively different behavior, with PE and SE approaching their saturation values according to power-law decay:
\be \label{eq:PE_SE_MFIM}
\mathcal{S}_2=\mathcal{S}^{\infty}_2-a_d t^{-\beta_S} \quad \mathrm{and} \quad \mathcal{M}_2= \mathcal{M}^{\infty}_2 -a^{\beta}_d t^{-\beta_M}
\ee
The saturation values $\mathcal{S}^{\infty}_2$ and $\mathcal{M}^{\infty}_2$ deviate from the Haar state averages, and this discrepancy grows with increasing $N$. Additionally, the timescale for equilibration of PE and SE in MFIM scales linearly with $N$.

\begin{figure}
    \centering
    \includegraphics[width=0.49\linewidth]{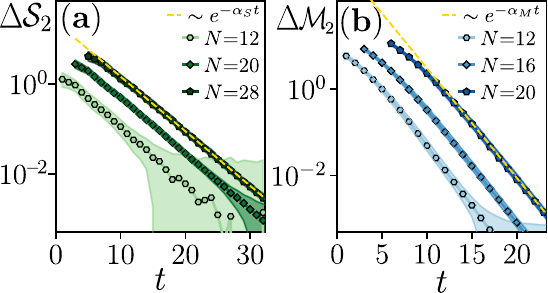}
    \caption{
Initial state dependence of the saturation of PE, (a), and SE, (b), in KIM. The 
averaged  $\Delta \mathcal{S}_2$ and $\Delta M_2$ decrease exponentially in time $t$, and are well fitted by $e^{-\alpha_{S,M} t}$, as discussed in the Main Text. The shaded regions represent the spread of $\Delta \mathcal{S}_2$ and $\Delta M_2$ with the variation of the initial state $\ket{\Psi_0}$, \eqref{eq:initial_state}. The lower/upper boundary of the shaded regions at fixed $t$ corresponds to $20$th/$80$th percentile of the distribution of $\Delta \mathcal{S}_2(t)$ and $\Delta \mathcal{M}_2(t)$, respectively shown in (a) and (b). The data for $N=20, 28$ in (a) and for $N=16,20$ in (b) have been shifted horizontally by factors of $4$
and $8$ for clarity of the plot.}
    \label{fig:EM1}
\end{figure}

\begin{figure}
    \centering
    \includegraphics[width=1\linewidth]{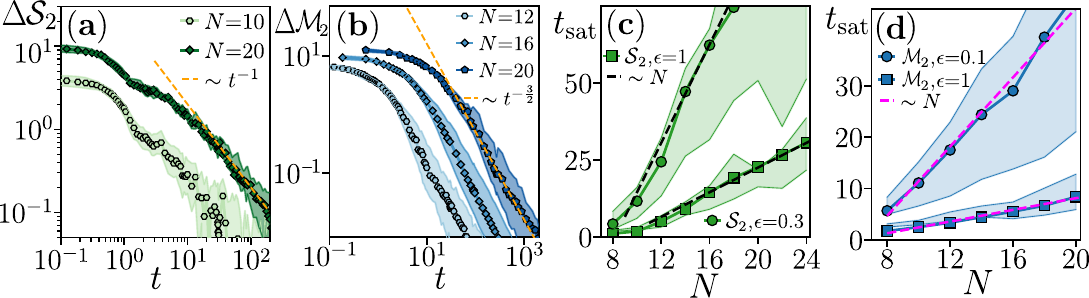}
    \caption{
Initial state dependence of the saturation of PE, (a), and SE, (b), in MFIM. The shaded regions in (a) show the 33rd/66th percentile of the distribution of  $\Delta_{ \mathcal{M}_2}(t)$, while, in (b), the shaded region boundaries correspond to $20$th/$80$th percentile of the distribution of $\Delta_{ \mathcal{S}_2}(t)$. (c) and (d) show the saturation time scales $t^{(\mathcal{S}_2,\mathcal{M}_2)}_{\mathrm{sat}}$ corresponding to the shaded regions of (a) and (b).  
  The data for $N=20$ in (a) and for $N=16,20$ in (b) have been shifted horizontally by multiplying by constant factors for clarity of the plot.
}
    \label{fig:EM2}
\end{figure}

In the following, we conduct an analysis of the PE and SE dynamics, focusing on their dependence on the choice of initial states. Figures~\ref{fig:EM1} and \ref{fig:EM2} illustrate our findings.

In Fig.~\ref{fig:EM1}, we focus on the behavior PE and SE in KIM dynamics. Panels (a) and (b) depict the exponential decay of the average deviations of $\Delta \mathcal{S}_2$ and $\Delta \mathcal{M}_2$ from the long-time saturation values $\mathcal{S}^{\infty}_2$ and $\mathcal{M}^{\infty}_2$ across more than $200$ randomly sampled initial 
states $\ket{\Psi_0}$ defined in Eq.~\ref{eq:initial_state}. These deviations are well-fitted by $e^{-\alpha_S t}$ and $e^{-\alpha_M t}$ respectively. The shaded regions indicate the spread of $\Delta \mathcal{S}_2$ and $\Delta \mathcal{M}_2$ due to variation in the initial state, with the spread decreasing as the system size $N$ increases. The decreasing spread shows that the saturation times $t^{(\mathcal{S}_2)}_{\mathrm{sat}}$, $t^{(\mathcal{M}_2)}_{\mathrm{sat}}$ discussed in the Main Text are well defined, and, for sufficiently large $N$, are independent of the choice of the initial state $\ket{\Psi_0}$.

In Fig.~\ref{fig:EM2}, we show analogous results for dynamics generated by the MFIM. Panels (a) and (b) demonstrate the power-law decay of the deviations $\Delta \mathcal{S}_2$ and $\Delta \mathcal{M}_2$, following $t^{-1}$ and $t^{-3/2}$ respectively. The shaded regions represent the spread of these deviations across the initial states, which only slightly diminishes with increasing $N$. Consequently, the saturation times $t'^{(\mathcal{S}_2)}_{\mathrm{sat}}$, $t'^{(\mathcal{M}_2)}_{\mathrm{sat}}$  for \textit{fixed initial state}, defined as times for which $\Delta \mathcal{S}_2(t)<\epsilon$ for a given initial state, are distributed over an interval which does not shrink with increasing $N$. The time scales $t^{(\mathcal{S}_2)}_{\mathrm{sat}}$, $t^{(\mathcal{M}_2)}_{\mathrm{sat}}$ presented in Fig.~\ref{fig:MFIM}~(e) of the Main Text, obtained from the averaged  $\Delta \mathcal{S}_2$ and $\Delta \mathcal{M}_2$, scale linearly with $N$, in the same way as analogous saturation times corresponding to \textit{typical}  $\Delta \mathcal{S}_2$ and $\Delta \mathcal{M}_2$ (data not shown). 
Panels (c) and (d) show the scaling of the saturation times with system size $N$, with shades corresponding to fixed lower and higher percentiles of the distribution of $\Delta \mathcal{S}_2(t)$ and  $\Delta \mathcal{M}_2(t)$. While there is a significant spread of these time scales, we observe an increase of the saturation times corresponding to the fixed lower and higher percentiles, corroborating our conclusions about the PE and SE saturation under the dynamics of MFIM.

The results discussed here underscore the important role of the initial state in the dynamics of PE and SE, with this dependence being more pronounced in MFIM than in KIM. For the Floquet dynamics, the saturation of PE and SE, up to a fixed tolerance $\epsilon$ occurs at a well defined time scale, independent of the system size $N$ (provided that $N$ is large enough). In contrast, the specific choice of initial state can significantly affect the time scale at which PE and SE saturate under the dynamics of MFIM. In spite of this spread, the saturation times in the latter case are longer, with characteristic time scales increasing linearly with system size.

\section{Disordered systems}
To further test the robustness of our conclusions, we include additional numerical results for both the KIM and the MFIM with on-site disorder. To that end, we modify the on-site fields in the Floquet operator of KIM and the Hamiltonian of MFIM from $h_i = (\sqrt{5}+1)/4$ to $h_i = (\sqrt{5}+1)/4 + \delta_i$, where $\delta_i$ are random numbers chosen independently with uniform probability in the interval $[-W, W]$.
\begin{figure}
    \centering
\includegraphics[width=0.76\linewidth]{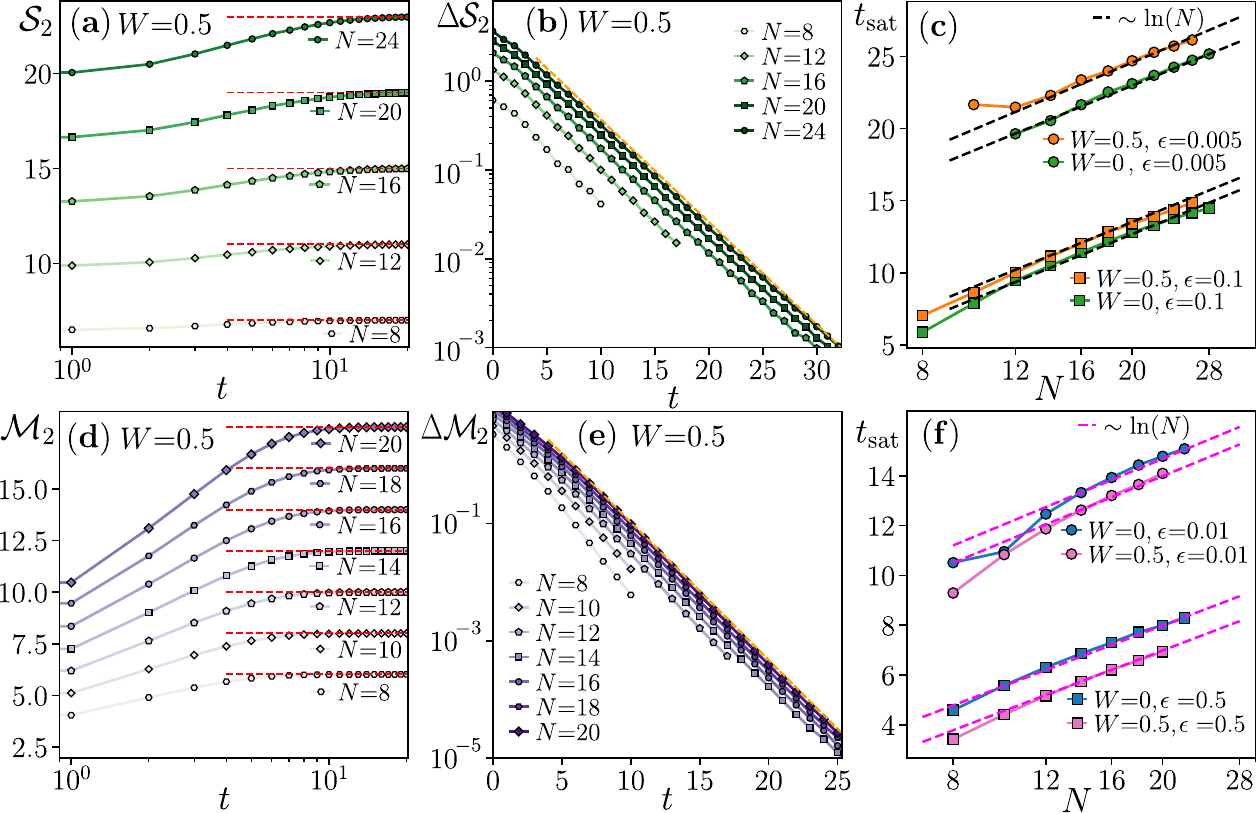}    
\caption{ {
Participation entropy, $\mathcal{S}_2$, and stabilizer R\'{e}nyi entropy, $\mathcal{M}_2$, under the dynamics of the KIM with on-site disorder of strength $W$. The phenomenology of PE and SE growth in the presence of disorder ($W = 0.5$) closely parallels that of the clean model ($W = 0$), as shown in panels (a), (b), (d), and (e). The saturation time $t_{\mathrm{sat}}$ is affected by the introduction of disorder, but the overall trend $t_{\mathrm{sat}} \propto \ln(N)$ remains unchanged.
} }
    \label{fig:fig4}
\end{figure}
\begin{figure}
    \centering
\includegraphics[width=0.76\linewidth]{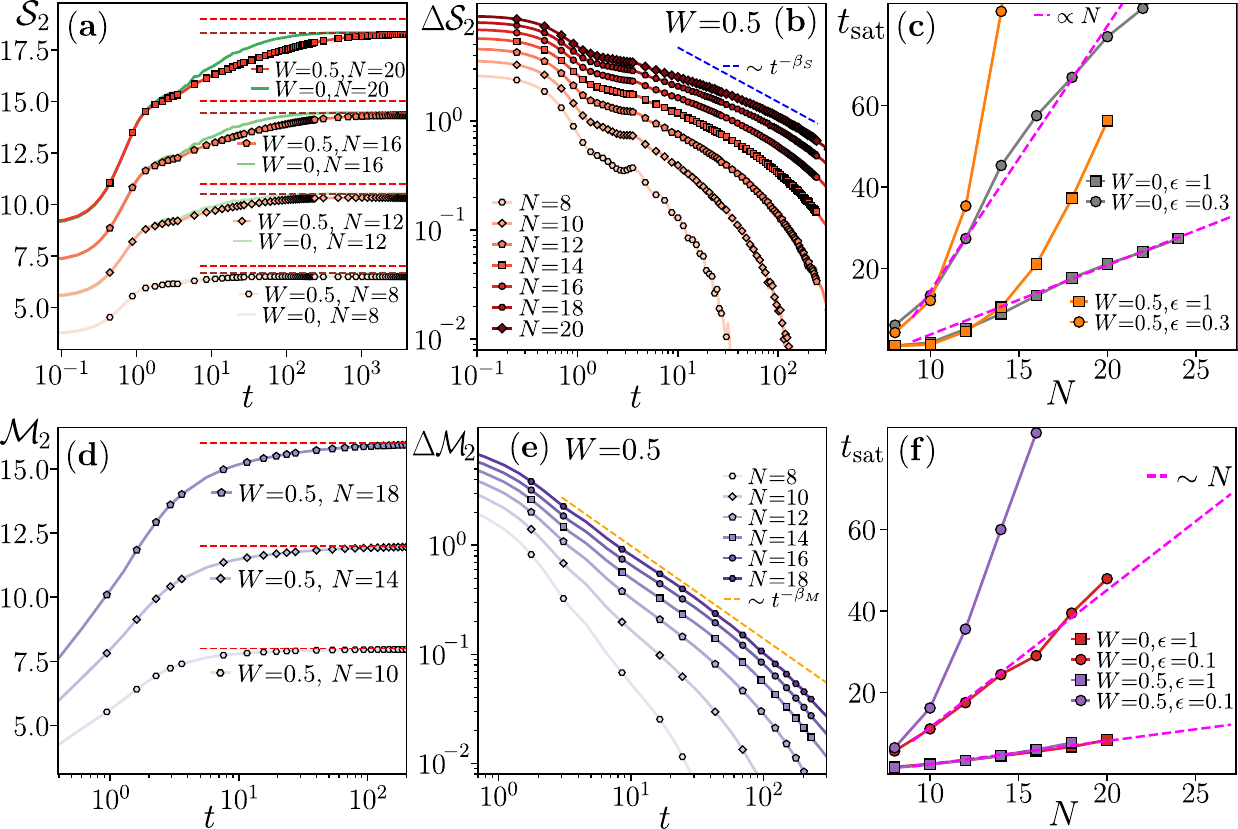}    
\caption{ {
Participation entropy, $\mathcal{S}_2$, and stabilizer R\'{e}nyi entropy, $\mathcal{M}_2$, under the dynamics of the MFIM with on-site disorder of strength $W$. The phenomenology of PE and SE growth in the presence of disorder ($W = 0.5$) closely parallels that of the clean model ($W = 0$), as shown in panels (a), (b), (d), and (e). The saturation time $t_{\mathrm{sat}}$ increases with the introduction of disorder, but the overall trend of (super)linear growth of $t_{\mathrm{sat}}$ with system size $N$ remains unchanged.
} }
    \label{fig:fig5}
\end{figure}

The results for PE and SE growth in the disordered KIM and MFIM, shown in Fig.~\ref{fig:fig4} and Fig.~\ref{fig:fig5}, confirm the robustness of the findings reported in the Main Text. The introduction of disorder with strength $W = 0.5$ does not alter the exponential (algebraic) saturation of PE and SE in the KIM (MFIM). Consequently, while the saturation times $t_{\mathrm{sat}}$ change parametrically, the stark contrast between the fast saturation ($t_{\mathrm{sat}} \propto \ln(N)$) in the KIM and the slower saturation of PE and SE in the MFIM persists even in the presence of disorder. 

In passing, we note that the saturation times in the MFIM for $W = 0.5$ appear to follow a superlinear scaling, which may be linked to diffusive energy transport in this model. However, the available numerical data do not allow us to draw a definitive conclusion.
Nevertheless, the clear difference in the PE and SE dynamics in the disordered KIM and MFIM confirms that the presence of energy conservation remains the decisive factor affecting the phenomenology of anticoncentration and magic spreading in quantum many-body systems. 
We note that these findings remain valid as long as the disorder is not too strong and the dynamics is ergodic. 
In contrast, the introduction of strong disorder leads to a \emph{breakdown of ergodicity} through the phenomenon of many-body localization (MBL), resulting in a distinct and richer phenomenology of PE and SE growth compared to ergodic systems, investigated recently in~\cite{Falcao25magic}.

\section{PE and SE in highly excited states}
Eigenvalues and eigenstates of the Hamiltonian determine the dynamics of a quantum system. To shed light on the saturation values of PE and SE under dynamics of MFIM, we investigate PE and SE across the spectrum of $H_{\mathrm{MFIM}}$. 

While the SE of eigenstates near the middle of the spectrum in ergodic many-body systems is well modeled by featureless Haar-random states~\cite{turkeshi2023paulispectrummagictypical}, the ground state of a local Hamiltonian is a highly atypical state, typically following an area law for entanglement entropy. Consequently, the SE in many-body ground states is limited, as observed in multiple studies in recent years (see, e.g.,~\cite{haug2023quantifying}).

\begin{figure}
    \centering
\includegraphics[width=0.77\linewidth]{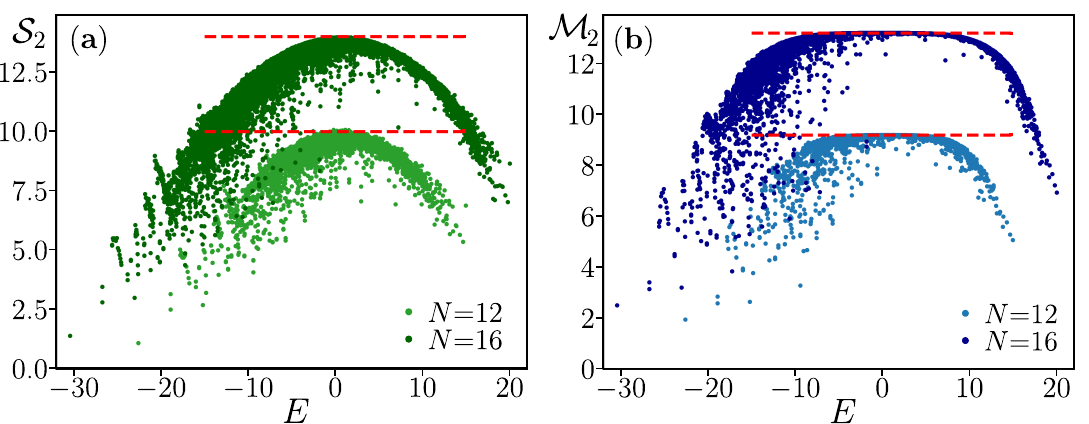}    
\caption{ \small{
Participation entropy, $\mathcal{S}_2$, and stabilizer R\'{e}nyi entropy, $\mathcal{M}_2$, in the eigenstates of the MFIM as functions of their energy $E$ for system size $N$. Eigenstates near the middle of the spectrum exhibit $\mathcal{S}_2$ and $\mathcal{M}_2$ values close to the Haar values, indicated by the red dashed lines. In contrast, the values of $\mathcal{S}_2$ and $\mathcal{M}_2$ are significantly suppressed for eigenstates with energies near the edges of the spectrum.
} }
    \label{fig:fig3}
\end{figure}
To bridge these two regimes, in Fig.~\ref{fig:fig3} we investigate the values of PE and SE across the spectrum of the MFIM. While the PE and SE of eigenstates in the middle of the spectrum approach the Haar values, both $\mathcal{S}_2$ and $\mathcal{M}_2$ are significantly reduced at low and high energy densities. This demonstrates the existence of subspaces in the Hilbert space, spanned by eigenstates of MFIM with limited $\mathcal{S}_2$ and $\mathcal{M}_2$, which restrict the effective volume of Hilbert space explored during the system's dynamics, resulting in the saturation values, $\mathcal{S}_2^{\infty}$ and $\mathcal{M}_2^{\infty}$ smaller than the Haar values.

\section{Dynamics of participation and stabilizer entropies at different $k$.}

The PE in the computational basis $\mathcal{B}_N = \{|\pmb{x}\rangle\}_{x=0}^{D-1}$ is defined as
\begin{equation} 
    \mathcal{S}_{k}(|\Psi\rangle) = \frac{1}{1-k} \log_2 \left[\sum_{\pmb{x} \in \mathcal{B}_N} p_{\pmb{x}}^k\right]\;, \quad p_{\pmb{x}} \equiv |\langle \pmb{x}|\Psi\rangle|^2\;, 
\end{equation}
where $p_{\pmb{x}}$ is the probability of finding the system in the basis state $\ket{\pmb{x}}$. The SE is defined as~\cite{leone2022stabilizerrenyientropy}
\begin{equation}
    \mathcal{M}_{k}(|\Psi\rangle) = \frac{1}{1-k} \log_2\left[\sum_{P \in \mathcal{P}_N} D^{k-1} \Xi_P^k\right]\;, \quad \Xi_P = \frac{\langle \Psi | P |\Psi\rangle^2}{D}\;, 
\end{equation}
with $\Xi_P$ denoting the Pauli spectrum~\cite{beverland2020lower,turkeshi2023paulispectrummagictypical}.
So far, we have focused on the growth of PE and SE under ergodic many-body dynamics for the R\'{e}nyi index $k = 2$. Here, we consider different values of $k$, highlighting the generality of the behavior reported in the Main Text.

\begin{figure*}
    \centering
    \includegraphics[width=1\linewidth]{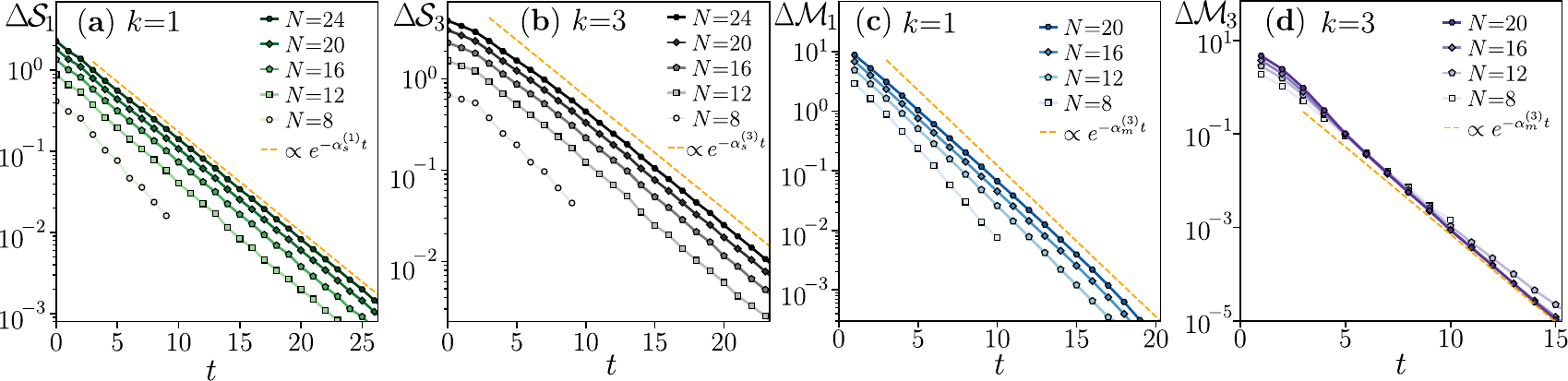}
    \caption{
    Saturation of the PE and SE for R\'{e}nyi index $k=1,3$ under the dynamics of the KIM. The difference $\Delta \mathcal{S}_k = \mathcal{S}^{\mathrm{Haar}}_k - \mathcal{S}_k(t)$ for $k = 1$ and $k = 3$, shown respectively in panels (a) and (b), decreases exponentially in time $t$ as $\Delta \mathcal{S}_k \propto e^{-\alpha^{(k)}_S t}$, with $\alpha^{(1)}_S \approx \alpha^{(3)}_S \approx 0.28(3)$. 
    The difference $\Delta \mathcal{M}_k = \mathcal{M}^{\mathrm{Haar}}_k - \mathcal{M}_k(t)$ for $k = 1$ and $k = 3$, shown respectively in panels (c) and (d), follows $\Delta \mathcal{M}_k \propto e^{-\alpha^{(k)}_M t}$, with $\alpha^{(1)}_M = 0.58(4)$ and $\alpha^{(3)}_M = 0.86(8)$.
    }
    \label{fig:KIMdiffK}
\end{figure*}

\begin{figure*}
    \centering
    \includegraphics[width=1\linewidth]{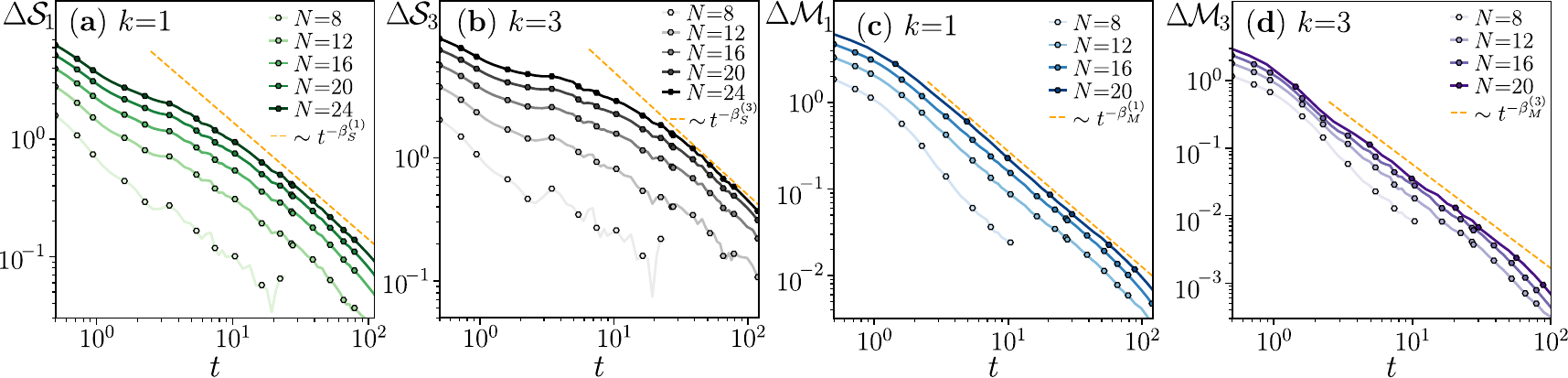}
    \caption{
    Saturation of the PE and SE for R\'{e}nyi index $k=1,3$ under the dynamics of the MFIM. The difference $\Delta \mathcal{S}_k = \mathcal{S}^{\infty}_k - \mathcal{S}_k(t)$ for $k = 1$ and $k = 3$, shown respectively in panels (a) and (b), decreases algebraically in time $t$ as $\Delta \mathcal{S}_k \propto t^{-\beta^{(k)}_S}$, with $\beta^{(1)} \approx \beta^{(3)} \approx 1 $. 
    The difference $\Delta \mathcal{M}_k = \mathcal{M}^{\infty}_k - \mathcal{M}_k(t)$ for $k = 1$ and $k = 3$, shown respectively in panels (c) and (d), follows $\Delta \mathcal{M}_k \propto t^{-\beta^{(k)}_M}$, with $\beta^{(1)}_M = 1.34(9)$ and $\beta^{(3)}_M = 1.5(2)$.
    }
    \label{fig:KMFIMdiffK}
\end{figure*}

Our results for $k = 1$ and $k = 3$ are shown in Fig.~\ref{fig:KIMdiffK} and Fig.~\ref{fig:KMFIMdiffK}, for the KIM and MFIM, respectively. The behavior for $k = 1$ and $k = 3$ faithfully follows the general trends reported in the Main Text for $k = 2$. In particular, both the PE and SE become extensive already at times of order unity, $t = O(1)$, and saturate to their asymptotic values $\mathcal{S}_k^{\infty}$ and $\mathcal{M}_k^{\infty}$ at longer times.

For the KIM, the results at $k = 1$ and $k = 3$ follow the phenomenology of random circuits~\cite{turkeshi2024hilbert, turkeshi2024magic}, with exponential saturation of $\mathcal{S}_k(t)$ and $\mathcal{M}_k(t)$ to the corresponding Haar values ($\mathcal{S}_k^{\infty} = \mathcal{S}^{\mathrm{Haar}}_k$ and $\mathcal{M}_k^{\infty} = \mathcal{M}^{\mathrm{Haar}}_k$). For $k = 3$, and more generally for any $k > 2$, the SE at long times is dominated by the identity Pauli string~\cite{turkeshi2023paulispectrummagictypical}, which decreases the timescale at which $\mathcal{M}_k(t)$ saturates to its long-time value—consistent with findings for random circuits~\cite{turkeshi2024magic}.

The dynamics of PE and SE for the MFIM at $k = 1$ and $k = 3$ are fully analogous to the results for $k = 2$ reported in the Main Text. In particular, our main conclusion regarding the stark contrast in PE and SE saturation time scales between Floquet and Hamiltonian systems remains valid for other values of the R\'{e}nyi index $k$.

\end{document}